\RequirePackage{fix-cm}
\documentclass[twocolumn]{svjour3}          
\smartqed  
\usepackage{graphicx}
\usepackage{indentfirst}
\usepackage{amsmath}
\usepackage{amsfonts}
\begin{document}

\title{Hard and soft excitation of oscillations in memristor-based oscillators with a line of equilibria
}


\author{Ivan A. Korneev         \and
        Tatiana E. Vadivasova \and
        Vladimir V. Semenov 
}


\institute{
I.A. Korneev \at 
Saratov State University, Astrakhanskaya str., 83, 410012, Saratov, Russia \\
              Tel.: +7-8452-210-710\\
              \email{ivan.korneew@yandex.ru}           
           	\and
T.E. Vadivasova \at 
Saratov State University, Astrakhanskaya str., 83, 410012, Saratov, Russia \\
              Tel.: +7-8452-210-710\\
              \email{vadivasovate@yandex.ru} 
             	\and
V.V. Semenov \at 
Saratov State University, Astrakhanskaya str., 83, 410012, Saratov, Russia \\
              Tel.: +7-8452-210-710\\
              \email{semenov.v.v.ssu@gmail.com}    
}

\date{Received: date / Accepted: date}

\maketitle

\begin{abstract}
A model of memristor-based Chua's oscillator is studied. The considered system has infinitely many equilibrium points, which build a line of equilibria. Bifurcational mechanisms of oscillation excitation are explored for different forms of nonlinearity. Hard and soft excitation scenarios have principally different nature. The hard excitation is determined by the memristor piecewise-smooth characteristic and is a result of a border-collision bifurcation. The soft excitation is caused by addition of a smooth nonlinear function and has distinctive features of the supercritical Andronov-Hopf bifurcation. Mechanisms of instability and amplitude limitation are described for both two cases. Numerical modelling and theoretical analysis are combined with experiments on an electronic analog model of the system under study. The issues concerning physical realization of the dynamics of systems with a line of equilibria are considered. The question on whether oscillations in such systems can be classified as the self-sustained oscillations is raised.
\keywords{memristor \and memristor-based oscillators \and border-collision bifurcations \and line of equilibria \and bifurcational analysis \and analog experiment}
\PACS{05.10.-a \and 05.45.-a \and 84.30.-r}
\end{abstract}

\section{Introduction}
\label{intro}
There are well-known three fundamental two-terminal passive electronic circuit elements: the resistor, the capacitor, and the inductor. The fourth element called the memristor was postulated by Leon Chua \cite{chua1971} in 1971. Chua's memristor relates the transferred electrical charge, $q(t)$, and the magnetic flux linkage, $\varphi(t)$: $d\varphi=M\cdot dq$, whence it follows that $M=M(q)=\dfrac{d\varphi}{dq}$. By using the formulas $d\varphi=Udt$ and $dq=idt$ ($U$ is the voltage across the memristor, $i$ is the current passing through the memristor) the memristor current-voltage characteristic can be derived: $U=M(q)i$. It means that $M$ is the current-controlled resistance (memristance) and depends on the entire past history of $i(t)$:
\begin{equation}
M(q)=\dfrac{d\varphi}{dq}=\varphi '\left( \int\limits_{-\infty}^{t}{i(t)dt} \right).
\label{M(q)}
\end{equation}
In a similar way, it can be written $dq=W \cdot d\varphi$, and then $W=W(\varphi)=\dfrac{dq}{d\varphi}$ and $i=W(\varphi)U$. Therefore $W$ is the flux-controlled conductance (memductance) and is the function of the entire past history of $U(t)$:
 \begin{equation}
W(\varphi)=\dfrac{dq}{d\varphi}=q '\left( \int\limits_{-\infty}^{t}{U(t)dt} \right).
\label{W(phi)}
\end{equation}
The Eq. (\ref{M(q)}) describes the charge-controlled memristor, and the Eq. (\ref{W(phi)}) describes the flux-controlled one. Two forms (\ref{M(q)}) and (\ref{W(phi)}) are mutually equivalent. Choice of the function describing the memristor depends on specific of each considered problem. The fundamental memristor property becomes highlighted from (\ref{M(q)}) and (\ref{W(phi)}). It is the dependence of electrical characteristics on past history of functioning. Initially, the memristor was introduced as realization of a hypothesis of relationship between the electrical charge and the magnetic flux linkage. Then the term "memristor" has been extended to the conception of "memristive systems". According to the paper \cite{chua1976} a class of the memristive systems is identified by continuous functional dependence of characteristics at any time on previous states of the memristive system, and is described by the following equations:
 \begin{equation}
 y=g(\boldsymbol{z},x,t)x,\quad
\boldsymbol{\dot{z}}=\boldsymbol{f}(\boldsymbol{z},x,t),
\label{mem-sys}
\end{equation}
where $x$ is the input signal, $y$ is the response of the system, the vector $\boldsymbol{z}\in \mathbb {R}^{n}$ denotes the system state, $\boldsymbol{f}(\boldsymbol{z},x,t)$ is a continuous $n$-dimensional vector function, $g(\boldsymbol{z},x,t)$ is a continuous scalar function. It is assumed that the state equation  $\boldsymbol{\dot{z}}=\boldsymbol{f}(\boldsymbol{z},x,t)$ has a unique solution for any initial state $\boldsymbol{z_{0}}\in \mathbb {R}^{n}$. It is important to note that such definition is mathematical interpretation, which does not concern a physical sense of the dynamical variables and their functional dependence. L.Chua and co-authors marked the fact that the Eqs. (\ref{mem-sys}) do not impose limitations on nature of the memristive systems \cite{ventra2009,chua2011}. For example, models of an electronic element thermistor and a discharge tube satisfy the definition (\ref{mem-sys}) of the memristive system \cite{chua1976,chua2013}. There are well-known examples of the memristive systems, which were developed more than a century ago \cite{gandhi2013}. As will be shown below, the memristive systems are of a frequent occurrence in different fields of physics as well as in neurodynamics and other sciences concerning biological systems. 

At the moment of publication the first Chua's papers concerning the memristor were less famous than now. A flurry of interest arose when scientists from the Hewlett Packard Company announced fabrication of TiO$_{2}$ oxides-based electronic two-terminal device, which exhibits features of the memristor dynamics \cite{strukov2008}. The authors have developed a mathematical model of the experimental prototype, which satisfies the definition (\ref{mem-sys}). The resistance of the experimental model presented in  \cite{strukov2008} depends on quantity of electricity (charge) passed through the device. In this context, the created passive two-terminal element can be assumed to be experimental realization of the memristor. Nevertheless, experimental demonstration has not closed the question of the memristor existence\footnote{Five decades later a hypothesis of relationship between the electrical charge and the magnetic flux linkage remains to cause debates in the scientific community \cite{vongehr2015}. There are many opinions on whether how that relationship correlates with theoretical concept of an electromagnetic field based on equations of classical electrodynamics (Maxwell's equations). Transformation of the memristor idea from a physical object to a generalised mathematical model also generates discussions. Undoubtedly, formulation of generalised interdisciplinary approach has a high importance. However, one should understand that experimental realization of the memristor in its initial form (as connecting link between the electrical charge and the magnetic flux) has not been invented yet.}, because a magnetic field is not used during experimental prototype functioning \cite{vongehr2015}. It should be pointed out, that the described in \cite{strukov2008} effects were highlighted during the late 1960s and called the "resistive switchings" \cite{argall1968}.

At the present time many objects, which exhibit the memristor dynamics, are known. Their characteristics reflect information about pre-history of operation. They are called the memristors in the following. Some of them are the oxide-based memristors \cite{strukov2008,argall1968,beck2000,mikhaylov2015,dearnaley1970,sawa2008,yang2012,kim2014,strachan2013,wu2013,mehonic2012,jeong2010,lee2016}, the polymer memristors \cite{krieger2004,liu2014,demin2015,berzina2009,erokhina2015}, the ferroelectric memristors \cite{moll1963,chanthbouala2012}, the memristors, based on resonant tunneling diodes \cite{buot1994,yilmaz2012,yilmaz2013}, the spintronic memristors, operation principles of whose are based on magnetic phenomena \cite{wang2009,chanthbouala2011,pershin2008}. Certain electronic circuits of two-port devices are realization of the memristor \cite{muthuswamy2010,bao2014}.

The memristors have a wide range of practical application. First of all, they are considered as a memory element. Memristor peculiarities can be used for analog circuit development, such like programmable amplifiers and attenuators, adaptive filters, to name only a few. The memristor attracts worldwide attention due to its potential applications in construction of next generation computers \cite{johnson2008,tour2008,ventra2013,yang2013}. A memristor potential for using in electronics is shown in the papers \cite{pershin2010,shin2011,li2012,gao2013,ascoli2013} and in the books  \cite{iu2012,radwan2015,tetzlaff2014,adamatzky2014,vourkas2016} in details. 

Appearance of the memristors has given an impulse to progression in certain fields of neurodynamics \cite{adamatzky2014,kozma2012}. Particularly, an analogy between the memristor dynamics and the behavior of a neural cell synapse \cite{jo2010,pershin2010-2,williamson2013,li2013,serb2016} makes be possible to consider the memristor as an element of neuromorphic electronic system working on principles of brain functioning\footnote{It should be noted, that experimental researches in neuromorphic electronics were carry out before memristor creation. In that case the memristor is considered as one of the two-terminal elements with the effect of resistive switchings, which can be used for neuromorphic electronic circuit development.} \cite{sheridan2016,gaba2013,emelyanov2016}. An interesting example of that similarity has been mentioned in the paper \cite{ziegler2012}. The authors have revealed the ability of memristor-based circuits to adapt to external signals and compared the obtained results with Pavlov's experiments in conditioned and unconditioned reflexes formation. 

Now the issues addressing the memristor remain to be in progress. Some recently obtained results are combined into the books \cite{vaidyanathan2017,ielmini2016}, which include the issues of modelling and theoretical aspects as well as circuit implementations and practical applications. Novel papers provide an encouraging pathway toward neuromorphic computing \cite{wang2017,adam2017,hamdioui2017,sheridan2017}, particularly in fuzzy logic systems \cite{zhang2017}. 

In the context of nonlinear dynamics, the memristor attracts an interest as an element, intrinsic properties of whose can essentially change the dynamics of electronic oscillatory systems and are responsible for qualitatively new types of the behavior. There are examples of memristor-based chaotic oscillators \cite{itoh2008,buscarino2012,buscarino2013,pham2013,gambuzza2015-2} and  Hamiltonian systems including the memristor \cite{itoh2011}. There is a variety of publications addressing the issue of the collective dynamics of coupled through the memristor oscillators and ensembles of coupled memristor-based oscillators: from synchronization of two chaotic or regular self-sustained oscillators coupled through the memristor \cite{frasca2014,volos2015,volos2016,frasca2015,ignatov2016} to
spatio-temporal phenomena in ensembles of interacting oscillators with adaptive coupling realized by the memristor or in ensembles of coupled memristor-based oscillators \cite{pham2012,gambuzza2015,buscarino2016,zhao2015}. Next interesting effect is a stochastic resonance, which has been revealed in the single memristor under the summary influence of noise and periodic signal \cite{stotland2012}. A broad variety of the effects is complemented by the existence of hidden attractors in models including memristor \cite{pham2015,chen2015,chen2015-2}.

Memristor-based systems can have manifolds of equilibria \cite{messias2010,botta2011,riaza2012,itoh2008}. They are the m-dimensional manifolds which consist of non-isolated equilibrium points. In the simplest case this manifold can exist as a line of equilibria (m = 1). Normally hyperbolic manifolds of equilibria are distinguished and their points are characterized by m purely imagine or zero eigenvalues, whereas all the other eigenvalues have nonzero real parts. In terms of the dynamical system theory, the systems whose phase space includes normally hyperbolic manifolds of equilibria can be referred to a special kind of systems with unusual characteristics. This class of dynamical systems has been considered in mathematical works, for example, in \cite{fiedler2000-1,fiedler2000-2,fiedler2000-3,liebscher2015,riaza2012,riaza2016}. It has been shown that their significant feature is the existence of so-called bifurcations without parameters, i.e., the bifurcations corresponding to fixed parameters when the condition of normal hyperbolicity is violated at some points of the manifold of equilibria. The existence of hidden attractors constitutes the second feature of such systems. According to the classification given in \cite{leonov2011,leonov2012,leonov2013,dudkowski2016}, all attractors can be divided into self-excited attractors and hidden attractors. A self-excited attractor has the basin of attraction which incloses the vicinity of an unstable equilibrium point. Alternatively, the basin of attraction of a hidden attractor does not intersect with the close neighborhood of all equilibrium point. It has been highlighted that hidden periodic and chaotic attractors are typical for the systems without equilibrium points or having only stable ones. In addition, it has been established in \cite{jafari2013,pham2016,pham2016-2} that all chaotic attractors are also hidden in the systems with a line of equilibria.

The self-oscillatory systems with a line of equilibria including the memristor were studied numerically and analytically in \cite{messias2010,botta2011}. A bifurcation of the periodic motion generation has been observed in such systems for fixed parameters. Both a set of stable equilibrium points and a set of non-isolated closed curves with different sizes can be observed depending on the initial conditions. All these trajectories create the system attractor which consists of some two-dimensional surface and stable branches of the line of equilibria. Peculiarities of the dynamics of the memristor-based oscillators with a line of equilibria were numerically and experimentally studied in the paper \cite{semenov2015} on an example of the system offered in \cite{itoh2008}. However, the paper \cite{semenov2015} is focused on study of self-oscillatory regimes in the presence of noise and their physical realization. In the present work we aim to rigorously reveal bifurcational mechanisms giving rise to formation of attractors, which consist of a continuous set of non-isolated closed curves and stable fixed points in the systems with a line of equilibria. We also present reasoning about classification of the oscillations corresponding to motion along the non-isolated closed curves.

\section{Models and methods}
\label{model}

\begin{figure}
\centering
\includegraphics[width=0.25\textwidth]{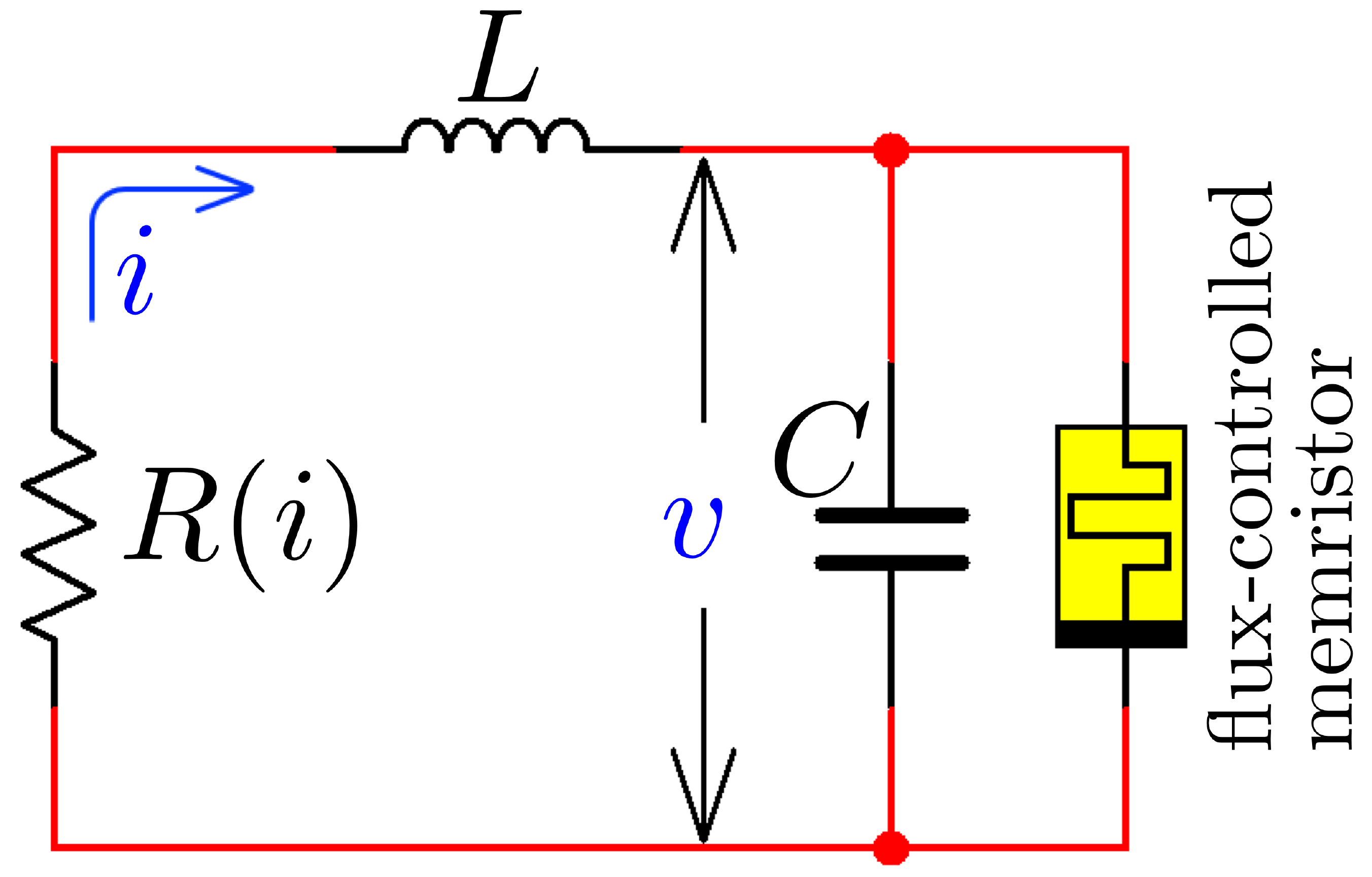} 
\caption{Schematic circuit diagram of the system under study.}
\label{fig1}
\end{figure}

Let us consider the model depicted in Fig. \ref{fig1}. This system was described in details in \cite{itoh2008}. It is the series oscillatory circuit including the element with negative resistance and the flux-controlled memristor. The element with negative resistance is assumed to be nonlinear in a general case, and its characteristics are described by the functional  dependence of the voltage on the current, $v_{R}(i)$. The presented in Fig. \ref{fig1} system is described by the following dynamical variables: $v$ is the voltage across the capacitance $C$,  $i$ is the current through the inductance $L$, and the magnetic flux, $\varphi$, controlling memristor. By using the Kirchhoff’s laws the following differential equations can be derived:
\begin{equation}
\label{physical_system}
\left\lbrace
\begin{array}{l}
i=C\dfrac{dv}{dt'}+W_{M}(\varphi)v, \\
L\dfrac{di}{dt'}+v_{R}(i)+v=0,\\
\dfrac{d\varphi}{dt'}=v,\\
\end{array}
\right.
\end{equation}
where $W_{M}(\varphi)$ is the flux-controlled conductance of the memristor. In the dimensionless variables $x=v / v_{0}$, $y=i/ i_{0}$ and $z=\varphi/(L\varphi_{0})$ with $v_{0}= 1$~V, $i_{0}=1$~A, $\varphi_{0}=\text{1 sec} \times v_{0}$ and the dimensionless time $t=[(v_{0}/(i_{0}L)]t'$, Eq.(\ref{physical_system}) can be re-written as,
\begin{equation}
\frac{dx}{dt}=\alpha (y-G_{M}(z)x), \frac{dy}{dt} = -x - f(y), \frac{dz}{dt}=x,
\label{initial_system}
\end{equation}
where $\alpha=(L/C)(i_{0}/v_{0})^{2}$  is the dimensionless parameter, which numerically equals to $L/C$. This parameter is assumed to be equal to unity in the following. The functions $f(y)$ and $G_{M}(z)$ are the dimensionless equivalent of the functions $v_{R}(i)$ and $W_{M}(\varphi)$ correspondingly. In the present work we consider the simplified model of the memristor introduced by L.Chua \cite{chua1971}. In the dimensionless form its characteristic is given by:
\begin{equation}
G_{M}(z)=
\begin{cases}
          a , & |z| < 1,\\
          b , & |z| \geq 1.
\end{cases}
\label{memristor}
\end{equation}

The variable $\varphi$ and its dimensionless analog $z$ play role of the memristor state variables. The memristor conductance, $W_{M}(\varphi)$ (or $G_{M}(z)$ in the dimensionless form), depends on these variables. Meanwhile, the instantaneous values of the variables $\varphi(t)$ and $z(t)$ are defined by entire past history of the $v(t)$ and $x(t)$ values. Indeed, meaning of the variables $\varphi$ and $z$ follows from the third equation of the systems (\ref{physical_system}) and (\ref{initial_system}):
\begin{equation}
\label{z-phi}
\varphi(t) = \int\limits_{-\infty}^{t}{v(t)dt}, \quad
z(t)=\int\limits_{-\infty}^{t}{x(t)dt}.
\end{equation}

\begin{figure}[t]
\centering
\begin{tabular}{cc}
\resizebox{0.8\columnwidth}{!}{\includegraphics{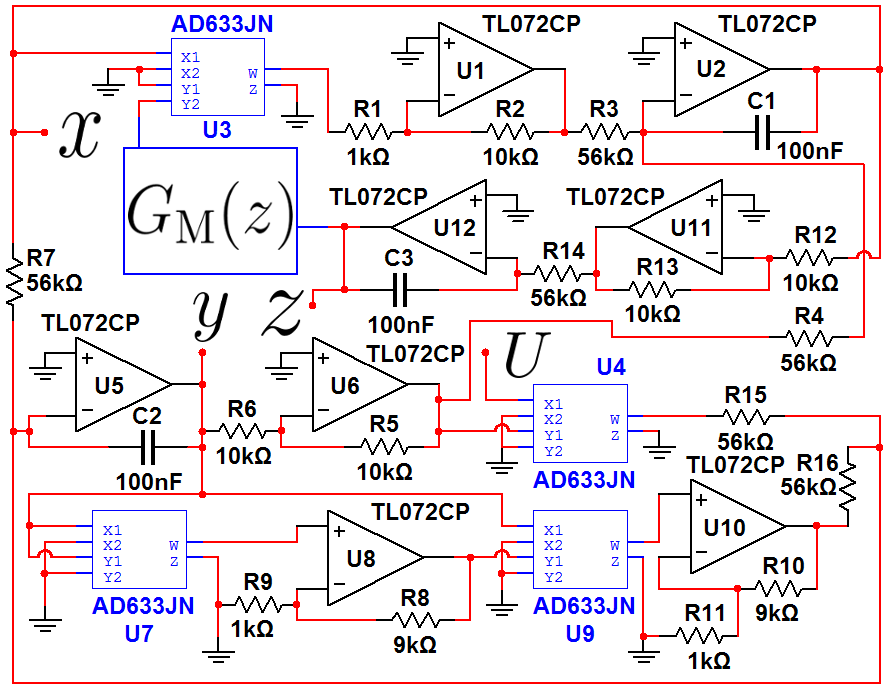} } \\
(a)\\
 \resizebox{0.78\columnwidth}{!}{\includegraphics{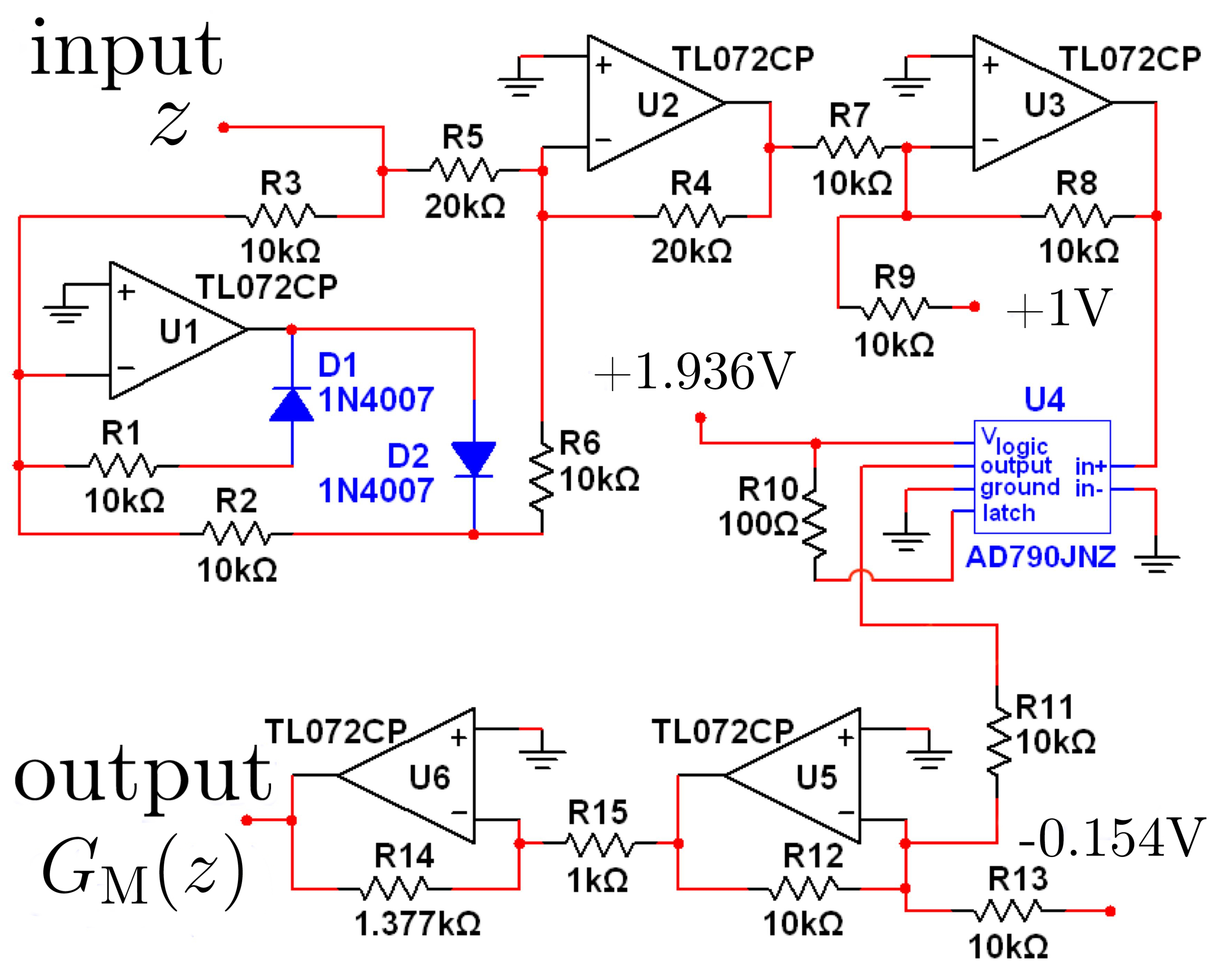} } \\
(b) \\
\end{tabular}
\caption{Schematic circuit diagram of the system (\ref{nonlinear}) analog model: (a) general circuit, (b) electronic realization of $G_{M}(z)$ function.}
\label{fig2}
\end{figure}

Two configurations of the studied system are considered. In the first case the oscillatory circuit includes the constant negative resistance $v_{R}(i)=-Ri$. Then the function $f(y)$ becomes $f(y)=-\beta y$ ($\beta$ is the fixed parameter) and the derived equations are:
\begin{equation}
\label{linear}
\left\lbrace
\begin{array}{l}
\dfrac{dx}{dt}=y-G_{M}(z)x, \\
\\
\dfrac{dy}{dt}=-x+\beta y, \\
\\
\dfrac{dz}{dt}=x,\\
G_{M}(z)=
\begin{cases}
          a , & |z| < 1,\\
          b , & |z| \geq 1,
\end{cases}
\end{array}
\right.
\end{equation}
where $\beta, a,b >0$ are the parameters (the parameters $a,b$ characterize the memristor, the parameter $\beta$ depends on properties of element with negative resistance). The second case provides the presence of the nonlinear element with the S-type current-voltage characteristic, which is approximated by the third-order polynomial: $v_{R}(i)=-k_{1}i+k_{3}i^{3}$. Then the function $f(y)$ of the system (\ref{initial_system}) is $f(y)=-\beta_{1}y+\beta_{3}y^{3}$ and the model equations become:
\begin{equation}
\label{nonlinear}
\left\lbrace
\begin{array}{l}
\dfrac{dx}{dt}=y-G_{M}(z)x, \\
\\
\dfrac{dy}{dt}=-x+\beta_{1} y-\beta_{3}y^{3}, \\
\\
\dfrac{dz}{dt}=x,\\
G_{M}(z)=
\begin{cases}
          a , & |z| < 1,\\
          b , & |z| \geq 1,
\end{cases}
\end{array}
\right.
\end{equation}
where $a,b,\beta_{1},\beta_{3} >0$ (the parameters $\beta_{1,3}$ are determined by the current-voltage characteristic of the nonlinear element).
In the present work the numerical and experimental studies of the system (\ref{linear}) carried out in the frameworks of the paper \cite{semenov2015} are complemented with theoretical analysis of bifurcations caused by parameter $\beta$ changing. The system (\ref{nonlinear}) is explored theoretically by using quasiharmonic reduction, and experimentally by means of both numerical and experimental modelling. Numerical simulations are carried out by integration of the Eqs. (\ref{linear}) and (\ref{nonlinear}) using the Heun method \cite{mannella2002} with time step  $t = 0.0001$ from different initial conditions. An experimental electronic setup, which is an analog model of the system (\ref{nonlinear}), was developed by using principles of analog modelling \cite{moss1989,luchinsky1998}. Motivation of that experimental approach choosing  comes from necessity to acquire instantaneous values of all three dynamical variables. It will be problematic to acquire time realization of the variable $z$ controlling the memristor in classical physical experiments using the  real memristor. The circuit diagram is shown in Fig. \ref{fig2}. The main part of analog models is an operational amplifier integrator, whose output voltage, $V_{out}$, is proportional to the input voltage, $V_{in}$, integrated over time: $V_{out}=-\dfrac{1}{R_{0}C_{0}} \int\limits_0^t V_{in}dt\quad$ or  $\quad R_{0}C_{0}\dot{V}_{out}=- V_{in} $, where $C_{0}$ and $R_{0}$ are the capacity and the resistance of the capacitor and the resistor at the integrator.
The circuit diagram shown in Fig. \ref{fig2}a contains three integrators, U2, U5 and U12, whose output voltages are taken as the dynamical variables, $x$, $y$ and $z$ respectively. Then the signals $x$, $y$ and $z$ are transformed in order to realize expressions of the right-hand side of Eqs. (\ref{nonlinear}). The necessary signal transformations are carried out by using the analog multipliers AD633JN and the operational amplifiers TL072CP connected in the inverting and noninverting amplifier configurations. Electronic circuit realization of the function $G_{M}(z)$ (see Fig. \ref{fig2}b) includes the fast precise voltage comparator AD790JNZ. Finally, transformed signals come to the input of the integrators as $V_{in}$. Time series are recorded from corresponding outputs (marked in Fig. \ref{fig2}a) using an acquisition board (National Instruments NI-PCI 6133). All signals were digitized at the sampling frequency of 50 kHz. 50 s long realizations were used for further offline processing. The circuit in Fig. \ref{fig2}a is described by the following equations:

\begin{equation}
\label{nonlinear_exp}
\left\lbrace
\begin{array}{l}
R_{0}C_{0}\dfrac{dx}{dt}=y-G_{M}(z)x, \\
\\
R_{0}C_{0}\dfrac{dy}{dt}=-x+\beta_{1} y-\beta_{3}y^{3}, \\
\\
R_{0}C_{0}\dfrac{dz}{dt}=x,\\
G_{M}(z)=
\begin{cases}
          a , & |z| < 1,\\
          b , & |z| \geq 1,
\end{cases}
\end{array}
\right.
\end{equation}
where $R_{0}=56k\Omega$, $C_{0}=100nF$ are the resistance and the capacity at the integrators, the parameter $\beta_{1}$ is proportional to the input voltage $U$ at the multiplier U4 (see Fig. \ref{fig2}a), $\beta_{1}=U/10$, the parameter $\beta_{3}$ is equal to unity, $a=0.02$, $b=2$. The equations (\ref{nonlinear_exp}) can be transformed to the dimensionless equations of the system (\ref{nonlinear}) by the following substitution: $t=t/\tau_0$, $x=x/v_{0}$, $y=y/v_{0}, z=z/v_{0}$, где $\tau_0=R_{0}C_{0}=5.6$~ms is the circuit's time constant, $v_{0}=1$~V.

\section{Circuit with constant negative resistance. Hard oscillation excitation}
\label{hard}
Let us consider the system (\ref{linear}) with fixed parameters $a=0.02$, $b=2$. This system has been described in the referred above papers \cite{itoh2008,botta2011,messias2010,semenov2015}. It is evident that the system (\ref{linear}) has a line of equilibria, i.e., each point on the axis OZ is an equilibrium point. One of the eigenvalues $\lambda_{i}$ of the equilibria is always equal to zero and the others depend on the parameters and the position of the point on the OZ-axis ($z$-coordinate):
\begin{equation}
\lambda_{1}=0, \\
\lambda_{2,3}=\frac{\beta-G_{M}(z)}{2} \pm \sqrt{\frac{(G_{M}(z)+\beta)^2}{4}-1}. 
\label{eigenvalues}
\end{equation}
Each point of the line of equilibria is neutrally stable in the OZ-axis direction. At the same time, for each point of the line of equilibria one can distinguish the plane $Q(x,y,z)$, which includes trajectories corresponding to attraction or repelling of this point in its vicinity. In the plane $Q(x,y,z)$ stability of the equilibrium point belonging to the line of equilibria is cracterized  by means of linear stability analysis used for isolated fixed points on the phase plane. Hereinafter, using the terms "stable" or "unstable" point in the line of equilibria, we mean the behavior of the trajectories in the neighbourhood of the equilibrium point in the plane $Q(x,y,z)$.

\begin{figure}[t]
\centering
\includegraphics[width=0.45\textwidth]{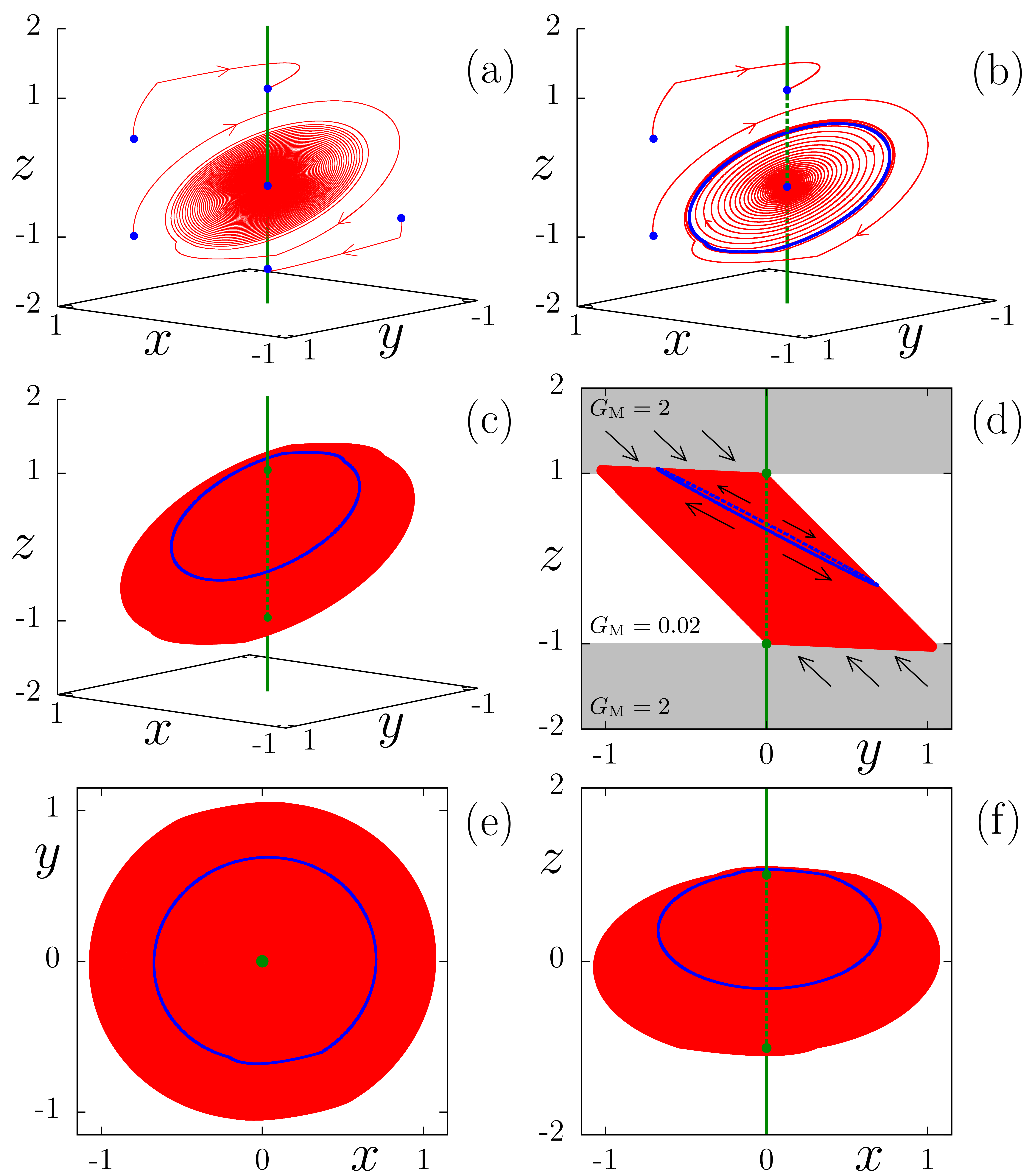} 
\caption{Trajectories and attractor in the phase space of the system (\ref{linear}): (a) motion to line of equlibria from different initial conditions  (trajectories are colored in red) for $\beta=0.01$; (b) motion to line of stable equilibria or to invariant closed curve (is colored in blue); (c) non-isolated closed curve, which corresponds to motion from initial condition ($x_{0}=y_{0}=0.5$, $z_{0}=-0.1$), on attractor; (d)-(f) Projections of attractor on the panel (c) and invariant curve in different planes. The panels (b)-(f) correspond to $\beta=0.05$. On all panels: a manifold of stable equilibria is shown by the gray solid line, a manifold of unstable equilibria is shown by the green dashed line, a two-dimensional surface formed by a manifold of invariant closed curves is colored in red (on the panels (c)-(f)), points corresponding to initial conditions are shown by blue filled circles. Other parameters are: $a=0.02$, $b=2$.}
\label{fig3}
\end{figure}

Increasing of the parameter $\beta$ from zero gives rise to the following bifurcational changes in the phase space. In case $0\le \beta<0.02$, an attractor of the system is a manifold of stable equilibria, and all trajectories are attracted to them [Fig. \ref{fig3}a]. When the parameter $\beta$ reaches the value $\beta=0.02$, equilibrium points with coordinate $-1<z<1$ become unstable, while other points of the line of equilibria with  $|z|\ge1$ remain to be stable (they become unstable at $\beta=2$). Starting from the vicinity of unstable point $(0;0;z\in(-1;1))$ the phase point moves away from the initial state and traces a spiral-like trajectory. That movement culminates in motion along an invariant closed curve (the blue line in Fig. \ref{fig3}b). Any changes of the initial conditions give rise to hit on another invariant closed curve. In that way, an attractor structure changes in the point $\beta=0.02$. After the bifurcation the attractor consists of a continuous set of closed curves which form a two-dimensional surface (like a whirligig) for $-1<z<1$ and of a set of points on the OZ-axis, for which $|z| \ge 1$. Further increasing of the parameter $\beta$ up to $\beta=2$ does not lead to qualitative and quantitative changes of the attractor.

Unfortunately, analysis of fixed point stability does not provide to uncover bifurcational mechanisms corresponding to formation of a two-dimensional surface for $-1<z<1$. In order to understand reasons of such an evolution, we transform the system (\ref{linear}) to the oscillatory form:
\begin{equation}
\label{linear-osc}
\left\lbrace
\begin{array}{l}
\dfrac{d^{2}x}{dt^{2}}=(\beta-G_{M}(z))\dfrac{dx}{dt}+(\beta G_{M}(z)-1)x, \\
\\
\dfrac{dz}{dt}=x. \\
\end{array}
\right.
\end{equation}
The first equation of the system (\ref{linear-osc}) does not include the variable $z$ in an explicit form. The dependence of the solution on the $z$ variable is defined by the function $G_{M}(z)$, which possesses two values. It makes be possible to solve the first equation of the system (\ref{linear-osc}) analytically and then to consider two cases of the $G_{M}(z)$ function. In a case of fixed $G_{M}(z)$ the first equation of the system (\ref{linear-osc}) is the linear oscillator with the frequency of oscillations $\omega_{0}=\sqrt{1-\beta G_{M}(z)}$. In that case the solution can be found as a harmonic function in complex representation: 
\begin{equation}
\label{x}
\left.
\begin{array}{l}
x(t)=Re \left\{ A(t)\exp{(i\omega_{0}t)}\right\}=\dfrac{1}{2}\{ A \exp{(i \omega_{0}t)} \\
+A^{*} \exp{(-i \omega_{0}t)} \} ,\\
\end{array}
\right.
\end{equation}
where $A(t)$ is the complex amplitude, $A^{*}(t)$ is the complex conjugate function. Then the first derivative is:
\begin{equation}
\label{dx}
\dfrac{dx}{dt}=\dfrac{1}{2}\left\{ i\omega_{0} A \exp {(i\omega_{0}t)} - i\omega_{0}A^{*} \exp {(-i\omega_{0}t)}    \right\} .
\end{equation}
The following condition for the first derivative is assumed to be satisfied: $\dfrac{dA}{dt} \exp{(i\omega_{0}t)}+\dfrac{dA^{*}}{dt} \exp{(-i\omega_{0}t)}=0$. With taking this condition into consideration the second derivative becomes:
\begin{equation}
\label{ddx}
\left.
\begin{array}{l}
\dfrac{d^{2}x}{dt^{2}}= \left( i\omega_{0} \dfrac{dA}{dt} - \dfrac{\omega_{0}^{2}A}{2} \right) \exp{(i\omega_{0}t)}\\
\\
-\left( i\omega_{0} \dfrac{dA^{*}}{dt} + \dfrac{\omega_{0}^{2}A^{*}}{2} \right) \exp{(-i\omega_{0}t)}.\\
\end{array}
\right.
\end{equation}
Then we substitute the Eqs. (\ref{x})-(\ref{ddx}) into the first equation of the system (\ref{linear-osc}):
\begin{equation}
\label{iwt}
\left.
\begin{array}{l}
i\omega_{0}\dfrac{dA}{dt}\exp{(i\omega_{0}t)}=\dfrac{(\beta-G_{M}(z))}{2}i\omega_{0}A\exp{(i\omega_{0}t)} \\
\\
-\dfrac{(\beta-G_{M}(z))}{2}i\omega_{0}A^{*}\exp{(-i\omega_{0}t)}\\
\\
-\omega_{0}^{2}A^{*}\exp{(-i\omega_{0}t)}.\\
\end{array}
\right.
\end{equation}
The coefficients at $\exp(i\omega_{0}t)$ in the left and right parts of the Eq. (\ref{iwt}) are equal. Using this equality, one can derive the equation for the complex amplitude $A$:
\begin{equation}
\label{linear_A}
\dfrac{dA}{dt}=\dfrac{\beta-G_{M}(z)}{2}A.
\end{equation}
The complex function $A(t)$ can be rewritten in polar coordinates:
 $A(t)=\rho (t) exp (i\phi(t))$. Then the system for a real amplitude and a real phase can be obtained:
\begin{equation}
\label{A-phi}
\left\lbrace
\begin{array}{l}
\dfrac{d\rho}{dt}=\dfrac{1}{2}(\beta-G_{M}(z))\rho, \\
\\
\dfrac{d\phi}{dt}=0. \\
\end{array}
\right.
\end{equation}
The equations of the system (\ref{A-phi}) are independent: the first equation does not include the phase, $\phi$, and the second equation does not include the amplitude, $\rho$. The phase, $\phi (t)$, is defined by initial value $\phi_{0}$. The problem concerning the existence of periodic oscillations and bifurcations in the system (\ref{linear-osc}) is reduced to amplitude equation (see system (\ref{A-phi})) analysis. The equlibrium solution of the amplitude equation is $\rho=0$, which corresponds to being in the point of equilibruim $x=0, y=0, z\in(-\infty;+\infty)$. The solution $\rho=0$ is stable for $\beta<G_{M}(z)$ and unstable for $\beta>G_{M}(z)$. Consideration of two cases $G_{M}(z)=0.02$ and $G_{M}=2$ gives rise to the conclusion written above: $\beta=0.02$ is the bifurcation value, which correspond to loss of stability of the equilibria $(x=y=0;z\in(-1;1))$ [Fig. \ref{fig3}c]. It can be inferred from the solution of the amplitude equation, that any stable or unstable solutions do not appear in the bifurcation moment. The phase point moves away from the equilibrium points, which lose the stability after the bifurcation. The system demonstrates harmonic oscillations with slowly increasing amplitude. The oscillations $x(t)$ can be described as $x(t)=\rho(t)\cos(\omega_{0}t+\phi_{0})$. It results from the first equation of the system (\ref{linear-osc}) that $\omega_{0}\approx 1$, because $\beta G_{M}(z)<<1$. Then it follows from the first equation of the system (\ref{linear}) that $y=\dfrac{dx}{dt}-G_{M}(z)x\approx \dfrac{dx}{dt}=-\rho(t)\sin(t+\phi_{0})$, because $G_{M}(z)=0.02$ in the subspace $-1<z<1$. The third equation of the system (\ref{linear}) allows us to obtain a formula for the $z(t)$-oscillations:
\begin{eqnarray}
\label{z}
\left.
\begin{array}{l}
z(t)=\int\limits_{0}^{t}\rho (t)\cos (t+\phi_{0}) dt \approx \\
 \rho(t) \int\limits_{0}^{t}\cos(t+\phi_{0})d(t+\phi_{0})=\\
  \rho (t)\sin (t+\phi_{0})+z_{0}.
\end{array}
\right.
\end{eqnarray}
Consequently, in case $0.02<\beta<2$ one can describe the oscillations of the dynamical variables of the system (\ref{linear}) for the initial conditions at the vicinity of points of equilibrium $(x=y=0;z\in(-1;1))$:
\begin{equation}
\label{xyz}
\left\lbrace
\begin{array}{l}
x(t)=\rho(t)\cos(t+\phi_{0}),\\
y(t)=-\rho(t)\sin(t+\phi_{0}),\\
z(t)=\rho(t)\sin(t+\phi_{0})+z_{0},
\end{array}
\right.
\end{equation}
where $z_{0}$ is the initial condition corresponding to each point in the line of equilibria. Numerical modelling of the oscillations resulting in motions along the invariant closed curve in planes ($x,y$), ($x,z$), ($y,z$) has confirmed a correctness of the solution (\ref{xyz}) (see Fig. \ref{fig3}d-f).

Amplitude growth of the oscillations $x(t)$ leads to increasing of the amplitudes of the $y(t)$ and $z(t)$ oscillations. It inevitably gives rise to coming of the dynamical variable $z(t)$ out of the range $[-1;1]$. Then the phase point 
reaches the subspace, where dissipation determined by memristor properties (for this regions $G_{M}(z)=2$, see the gray areas in Fig. \ref{fig3}d) exceeds energy pumping characterized by the negative resistance (the parameter $\beta$). The amplitude growth stops. As a result, oscillations with constant amplitude are observed. The absolute value of the dynamical variable $z$ reaches the neighborhood of unity once a period of the oscillations. 
These established oscillations correspond to motion along an invariant closed curve on the attractor. The described transition to the oscillatory regime is characterized by amplitude hopping and can be considered as hard excitation of the oscillations. In such a case the hard oscillation excitation is not associated with local bifurcations of limit sets. This scenario stands out from classical bifurcational mechanisms. For example, it is differ from the subcritical Andronov-Hopf bifurcation corresponding to abrupt self-sustained oscillation excitation after collision of an unstable limit cycle and a stable fixed point. In the system (\ref{linear}) bifurcational transition is determined by the memristor piecewise-smooth characteristic. 

The bifurcations caused by piecewise-smooth nonlinearity are well-known \cite{zhusubaliyev2003}. They are a frequent occurrence in mechanics \cite{leine2004}, biology\cite{dejong2004}, switching electronics \cite{dibernardo1998,banerjee2000}, power electronics \cite{banerjee2001}, to name only a few. Different kinds of such bifurcations are distinguished\footnote{Depending on authors, classification can slightly vary.}. The first one is represented by border-collision bifurcations (also known as C-bifurcations). They occur when the phase point crosses a boundary of discontinuity and reaches a region of the phase space where the dynamics of the system is different. Such bifurcations can be induced when limit sets cross a discontinuity border in the phase space and then bifurcate\footnote{Sometimes bifurcations of stationary states are distinguished and called  "boundary equilibrium bifurcations" \cite{dibernardo2010}}. This class of bifurcations has been studied on examples of discrete maps \cite{nusse1992,jain2003,banerjee2000,gardini2011} and continuous-time systems \cite{dibernardo1999,dibernardo2008,zhusubaliyev2001}. 
Grazing bifurcations represent the second class of discontinuity-induced bifurcation. These occur when the limit sets tangent a smooth boundary dividing distinct regions in phase space where the vector field is smooth \cite{dibernardo2001,dibernardo2001-2}. 

The described bifurcational change in the system (\ref{linear}) is a border-collision bifurcation. It is related with the situation, when a trajectory of periodic motion passes through either plane sewing the piecewise-smooth function $G_{M}(z)$.

\section{Circuit with nonlinear negative resistance. Soft oscillation excitation}
\label{soft}

\begin{figure}[t]
\centering
\includegraphics[width=0.45\textwidth]{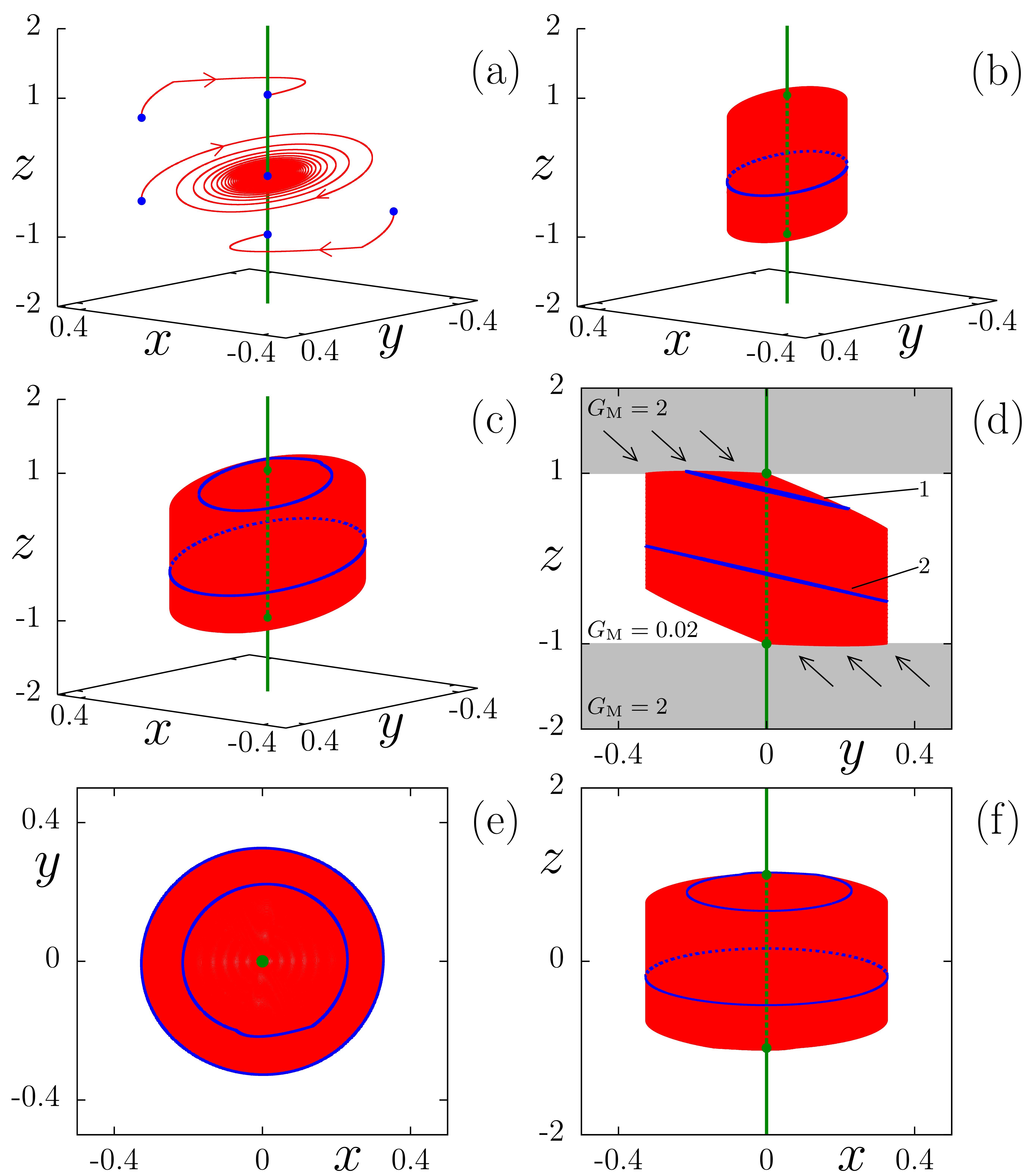} 
\caption{Phase space of the system (\ref{nonlinear}). (a) $\beta_{1}=0.01$. Trajectories in phase space (are colored in red) tracing motion to line of equilibria from different initial conditions; (b) Attractor corresponding to $\beta_{1}=0.05$; (c) Attractor corresponding to $\beta_{1}=0.1$; (d)-(f) Projections of attractor on panel (c) in planes ($y,z$),($x,y$),($x,z$).
On all panels: a manifold of stable equilibria is shown by the green solid line, a manifold of unstable equilibria is shown by the green dashed line, points corresponding to initial conditions are shown by blue filled circles. On the panels (b)-(f): A two-dimensional surface formed by a manifold of invariant closed curves is colored in red. Two invariant closed curves are colored in blue. Other parameters are: $a=0.02$, $b=2$, $\beta_{3}=1$.}
\label{fig4}
\end{figure}

Let us consider the system (\ref{nonlinear}) and fix the parameters $\beta_{3}=1$, $a=0.02$, $b=2$. The system has an infinite number of equilibrium points ($0;0;z\in(-\infty;+\infty)$) forming a line of equilibria in the phase space. Eigenvalues of the equilibria for the systems (\ref{linear}) and (\ref{nonlinear}) are the same (the formulas (\ref{eigenvalues})). The parameter value $\beta_{1}=0.02$ in the system (\ref{nonlinear}) is the bifurcational value corresponding to loss of stability of the equilibria $(x=y=0,z\in(-1;1))$ as well as in the system (\ref{linear}). After the bifurcation an attractor of the system (\ref{nonlinear}) consists of a continuous set of closed curves which form a two-dimensional surface and of a set of stable points of equilibrium on the OZ-axis. Bifurcational changes are illustrated in Fig. \ref{fig4}a,b. In contrast to the system (\ref{linear}), in this case smooth increasing of the parameter $\beta_{1}$ gives rise to gradual growth of the  amplitude of the $x(t)$, $y(t)$ and $z(t)$ oscillations corresponding to motions along invariant closed curves [Fig. \ref{fig4}b,c]. When the parameter $\beta_{1}$ growths, the two-dimensional surface formed by a manifold of invariant closed curves becomes expanded. Further increasing of the parameter $\beta_{1}$ results in formation of the attractor, which is identical to the attractor [Fig. \ref{fig3}c] of the system (\ref{linear}).

In order to understand reasons of the attractor transformation caused by the parameter $\beta_{1}$ growth, theoretical analysis by means of quasiharmonic reduction is carried out. The oscillatory form of the system (\ref{nonlinear}) is:
\begin{equation}
\label{nonlinear-osc}
\left\lbrace
\begin{array}{l}
\dfrac{d^{2}x}{dt^{2}}=-x-G_{M}(z)\dfrac{dx}{dt}+\beta_{1} \left( \dfrac{dx}{dt}+G_{M}(z)x\right) \\
-\left( \dfrac{dx}{dt}+G_{M}(z)x\right)^{3} , \\
\\
\dfrac{dz}{dt}=x. \\
\end{array}
\right.
\end{equation}
Periodic solution of the first equation of the system (\ref{nonlinear-osc}) on the frequency $\omega_{0}=\sqrt{1-\beta_{1}G_{M}(z)}$ is sought. Two values of the $G_{M}(z)$ are considered independently as well as in the linear system. 
Substituting Eqs. (\ref{x})-(\ref{ddx}) into the first equation of the system (\ref{nonlinear-osc}), we approximate all fast oscillating terms by their averages over one period $T=2\pi/\omega_{0}$ which gives zero. Then we obtain the equation for the complex amplitude, $A$:
\begin{equation}
\label{nonlinear_A}
\left.
\begin{array}{l}
\omega_{0}\dfrac{dA}{dt}=-\dfrac{i\left( \omega_{0}^{2}-1\right)}{2}A-\dfrac{G_{M}(z)\omega_{0}}{2}A\\
\\
-\dfrac{4\beta_{1}-3\left( G_{M}^{2}(z)+\omega_{0}^{2}\right) |A|^{2}}{8}\bigg( iG_{M}(z)-\omega_{0}\bigg) A.
\end{array}
\right.
\end{equation}
The Eq. (\ref{nonlinear_A}) can be rewritten as a system of equations for a real amplitude and a real phase:
\begin{equation}
\label{A-phi-nonlinear}
\left\lbrace
\begin{array}{l}
\dfrac{d\rho}{dt}=-\dfrac{G_{M}(z)}{2}\rho+\dfrac{4\beta_{1}-3\left( G_{M}^{2}(z)+\omega_{0}^{2}\right)\rho^{2}}{8}\rho, \\
\\
\dfrac{d\phi}{dt}=\dfrac{3\left( G_{M}^{2}(z)+\omega_{0}^{2}\right)\rho^{2}}{8\omega_{0}}G_{M}(z). \\
\end{array}
\right.
\end{equation}

As well as in the previous case, the problem of the periodic motion existence and bifurcations in the system (\ref{nonlinear-osc}) is reduced to amplitude equation analysis. The amplitude equation of the system (\ref{A-phi-nonlinear}) has two solutions. The first solution $\rho_{1}= 0$, corresponds to the equilibria ($x=0$, $y = 0$, $z \in(-\infty; +\infty)$), which are stable for $\beta_{1} < G_{M}(z)$ and unstable for $\beta_{1} > G_{M}(z)$. The second equilibrium solution, $\rho_{2}=\dfrac{2}{\sqrt{3}}\sqrt{\dfrac{\beta_{1}-G_{M}(z)}{G_{M}^{2}(z)+\omega_{0}^{2}}}$, appears at $\beta_{1}=G_{M}$, is stable and corresponds to invariant closed curve appearance in the vicinity of an equilibrium point. Considering two cases $G_{M}(z) = 0.02$ and $G_{M} = 2$ for subspaces $|z|<1$ and $|z|\ge1$, we obtain the following picture of bifurcational changes for increasing parameter $\beta_{1}$. For $\beta_{1}=0.02$ the stable equilibria $(x=y=0;|z|<1)$ become unstable, and invariant closed curves appear at their vicinity. The amplitude of the oscillations $x(t)$ corresponding to motions along the invariant closed curve increases gradually starting from zero with the parameter $\beta_{1}$ growth. The law of amplitude variation is: 
\begin{equation}
\label{pho-beta-theor}
\rho(\beta_{1})=\dfrac{2}{\sqrt{3}}\sqrt{\dfrac{\beta_{1}-0.02}{0.02^{2}+1}}\approx \dfrac{2}{\sqrt{3}}\sqrt{\beta_{1}-0.02}, 
\end{equation}
because $\omega_{0}\approx1$ in the neighborhood of the line of equilibria $x=y=0$ (it comes from the first equation of the system (\ref{nonlinear-osc})). Such scenario of oscillation excitation in the system (\ref{nonlinear}) is the soft mechanism. The performed analysis of system's (\ref{nonlinear}) dynamics by means of quasiharmonic reduction allows us to make clear the attractor structure after the bifurcation. The attractor includes two semiaxis of the stable equilibria $(x=y=0;|z|\ge1)$ and a closed surface which consists of several parts. The central part is formed by a continuous manifold of identical invariant closed curves (one of them is the closed curve 2 in Fig. \ref{fig4}c-f). The central surface corresponds to the region of space, where character of the oscillations is defined by properties of the nonlinear negative resistance, similarly to the Van der Pol self-sustained oscillator. In this region the instantaneous value $z(t)$ does not reach unity. The dependence of the oscillation $x(t)$ amplitude on the parameter $\beta_{1}$ (Eq. (\ref{pho-beta-theor})) was derived for the central surface. If the value $|z(t)|=1$ is reached once a period of oscillations (see for example the curve 1 in Fig. \ref{fig4}c-f), then limitation of an oscillation amplitude is realized by the combined impact of the nonlinear element and the memristor. The influence of the memristor is identical in the systems (\ref{linear}) and (\ref{nonlinear}): sharp increase of dissipation outside of the subspace $z\in[-1;1]$.

\section{Oscillations or self-sustained oscillations?}
\label{self-osc}
The described above soft mechanism of oscillation excitation seems to be similar to self-oscillatory regime realization in nonlinear dissipative systems, which is associated with the supercritical Andronov-Hopf bifurcation \cite{andronov1966}. Indeed, the soft excitation of the oscillations in the system (\ref{nonlinear}) consists in loss of stability in a certain part of the line of equilibria and generation of a manifold of invariant closed curves. Moreover, the dependence of the periodic solution amplitude as a square root of supercriticality (as in the system (\ref{nonlinear}), see Eq. (\ref{pho-beta-theor})) is typical for self-sustained oscillators, which realize the supercritical Andronov-Hopf bifurcation. However, full analogy is incorrect. Result of the Andronov-Hopf bifurcation in the nonlinear dissipative systems is transition to the regime of self-sustained oscillations, an image of which is a stable limit cycle, a closed isolated trajectory. Before the Andronov-Hopf bifurcation an attractor, a stable fixed point corresponding to absence of oscillations, exists in the phase space. After the bifurcation other attractor, a stable limit cycle, appears. However, invariant closed curves in the phase space of the systems (\ref{linear}) and (\ref{nonlinear}) are not attractors in themselves. One cannot distinguish a finite basin of attraction of each invariant curve. Similarly, each fixed point in the line of equilibria is not the attractor in itself, one of the eigenvalues of these points is always zero.

Because of these reasons, we do not use the term "self-sustained oscillations" for identification of motions along the invariant closed curve in the systems (\ref{linear}) and (\ref{nonlinear}). The term "self-sustained oscillations"  means the independence of the oscillatory characteristics on the initial conditions in a finite region of the phase space \cite{andronov1966}. That independence is realized in the system (\ref{nonlinear}) in the context of the $x(t)$ and $y(t)$ oscillations. The amplitude of the $x(t)$ and $y(t)$ oscillations is constant for a certain region of the phase space. However, the same cannot be said about the oscillations $z(t)$, the mean value of whose continuously depends on the initial conditions. The issues concerning classification of the oscillations in the systems with a line of equilibria require individual consideration and will be analyzed in further researches. 
\section{Analog experiment}
\begin{figure}[t]
\centering
\includegraphics[width=0.45\textwidth]{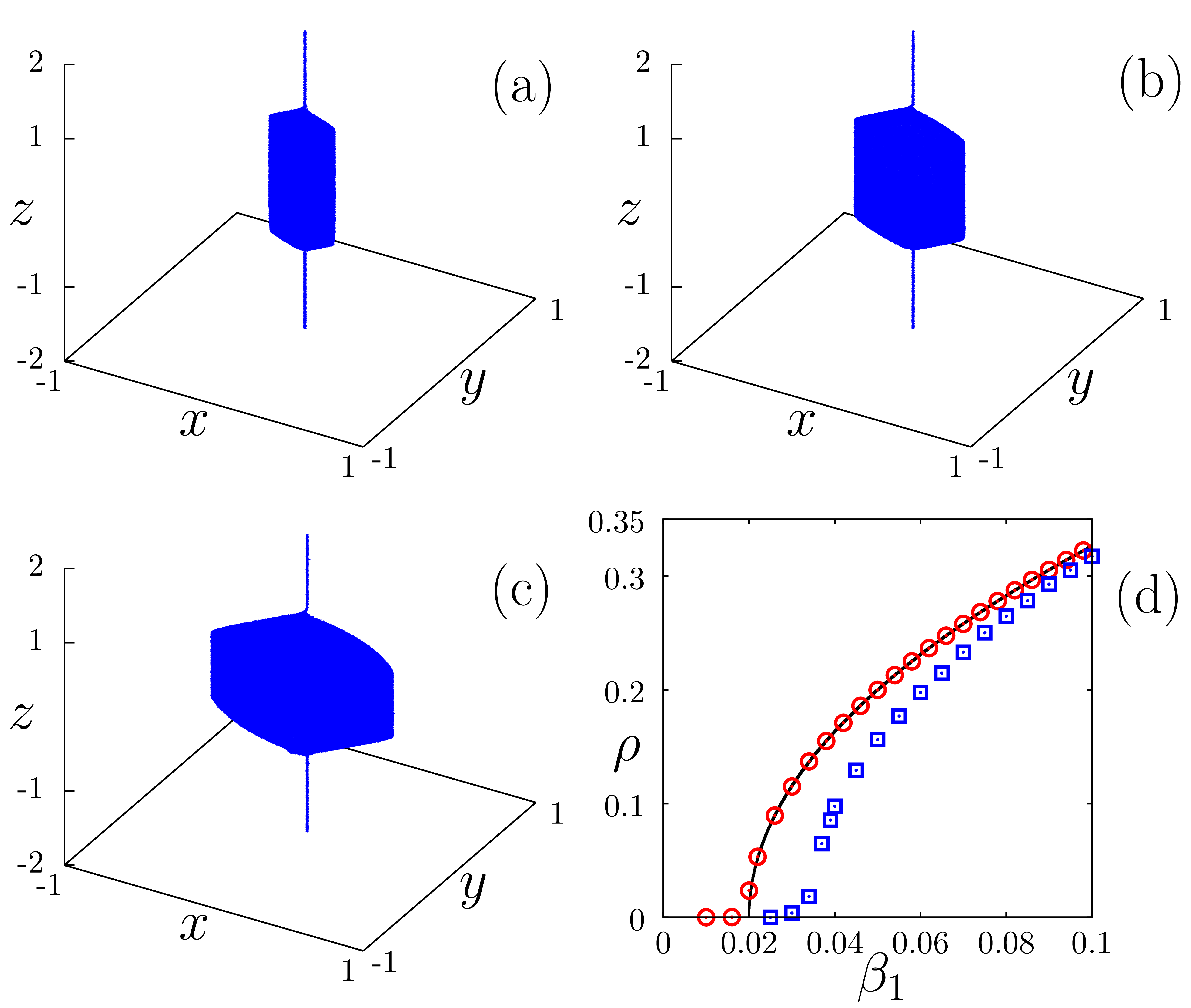} 
\caption{Trajectory traced by experimental setup (the system (\ref{nonlinear_exp})): $\beta_{1}=0.04$ (a); $\beta_{1}=0.1$ (b); $\beta_{1}=0.2$ (c). (d) Dependence of amplitude of oscillations $x(t)$ of the system (\ref{nonlinear}), for which $|z(t)|$ is less than unity, on parameter $\beta_{1}$: derived analytically (the solid black line) and registered in numerical modelling (the red circles). The dependence of the maximal amplitude of the oscillations  $x(t)$ during operation of the experimental setup (\ref{nonlinear_exp}) on the parameter $\beta_{1}$ is shown by blue squares.}
\label{fig5}
\end{figure}  
Experimental consideration of the system (\ref{linear}) carried out by means of analog modelling in \cite{semenov2015} has shown impossibility to observe stationary oscillations corresponding to motion along invariant closed curves. Since initial power-up moment the variable $z$ 
monotonically increases and reaches its maximal value limited by power-supply voltage. During this process oscillations of the experimental facility trace a half of the attractor (a two-dimensional semisurface and a line of equilibria) corresponding to positive $z$. This is due to the fact that equations describing experimental setup functioning do not take into consideration certain peculiarities of real operational amplifiers. For example, account of the small but nonzero current at the input of the operational amplifiers gives rise to the existence of an extremely small but nonzero constant in the third equation of the systems (\ref{linear}) and (\ref{nonlinear}). It has a principal importance, because the presence of a nonzero constant in the third equation $\dfrac{dz}{dt}=x$ destroys the line of equlibria. Analysis of equilibrium points in the modified systems (\ref{linear}) and (\ref{nonlinear}) with the third equation $\dfrac{dz}{dt}=x+\varepsilon$ highlights this fact. The numerically modelled dynamics of the system (\ref{nonlinear}) for nonzero $\varepsilon$ is in a good correspondence with the behavior of the experimental facility (\ref{nonlinear_exp}). Connecting an additive constant voltage to the integrator U12 input (see Fig.\ref{fig2}a), one can slow down and reverse monotonic motion along OZ-axis. This approach was used in \cite{semenov2015}. It is important to note that monotonic motion along OZ-axis can be slowed down and reversed but cannot be stopped. It is impossible to counterbalance with absolute accuracy the impact of the internal parameter $\varepsilon$ in real experiments. Therefore stationary oscillatory regimes cannot be obtained in analog experiment. Because of these reasons experimental investigation of the dynamics depending on the initial conditions is unachievable. Nevertheless, there is a similarity between the dynamical model and the experimental setup. During the process of $z$-coordinate monotonic drift (this process is limited by power-supply voltage) the experimentally acquired trajectory traces a figure, which is very similar to the mathematical model attractor corresponding to analogous parameter values [Fig. \ref{fig5}a-c]. 
 
The dynamics of the experimental setup with the parameter $\beta_{1}$ growth has shown structural change at $\beta_1\approx 0.03$. For $\beta_{1}>0.03$ the periodic oscillations $x(t)$ and $y(t)$ are induced when the $z(t)$ values are inside the range (-1;1) (see trajectories in Fig. \ref{fig5}a-c). The maximal amplitude of the $x(t)$ oscillations corresponding to passing through the area $z\in(-1;1)$ gradually increases with the parameter $\beta_1$ growth. The experimentally registered maximal amplitude of the $x(t)$ oscillations is close to the theoretically calculated amplitude of the oscillations $x(t)$ corresponding to motion along an invariant closed curve in the central part of a two-dimensional surface (like curve 2 in Fig. \ref{fig4}d) in the mathematical model phase space [Fig. \ref{fig5}d].

\section{Line of equilibria in context of memristor properties}
The significant feature of the memristor is a continuous dependence of characteristics on the all past history of functioning. In the context of dynamical systems it is associated with continuous dependence of the oscillatory dynamics properties on the initial conditions. The dynamics of systems with a line of equilibria fully reflects this peculiarity. When an attractor of the system is a line of equilibria (before the bifurcation of oscillation excitation), the phase point becomes attracted and reaches the line of equilibria in some point with coordinates ($0;0;z$). Change of the initial conditions gives rise to hit to another point of the line of equilibria ($0;0;z^{*}$). Continuous variation of the initial conditions leads to continuous change of $z$-coordinate of the point on the line of equilibria ending the motion to the line of equilibria. Continuous dependence on the initial conditions remains to exist after the bifurcation of oscillation excitation. Depending on the initial conditions, the phase point reaches one of the non-isolated invariant closed curves or on the point on the line of equilibria. 

There is the second important question. In the presence of fluctuations stochastic oscillations in a system with a line of equilibria are nonstationary and are similar to the behavior of a Brownian particle in the presence of random pushes only, i.e., we deal with a Wiener process (this issue was considered in \cite{semenov2015}). This fact correlates with the memristor properties. It has been shown in \cite{slipko2013,diventra2013} that memristor would erratically change its state in course of time just under the influence of noise. Therefore it can be concluded, that memristors whose characteristics depend solely on the current or voltage history would be unable to protect their memory states against unavoidable Johnson--Nyquist noise and permanently suffer from information loss, a so-called "stochastic catastrophe". 
\section*{Conclusions}
\label{conclusions}
Bifurcational mechanisms of the periodic solution appearance in systems with a line of equilibria have been analyzed on the example of the memristor-based system. The hard excitation scenario results from the piecewise smooth memristor characteristic and is associated with border-collision bifurcations. Soft excitation of the oscillations consists in the invariant closed curve emergence from become unstable equilibrium points belonging to the line of equilibria. The soft excitation of the oscillations has a similar character as compared to the supercritical Andronov-Hopf bifurcation corresponding to soft excitation of the oscillations in nonlinear dissipative systems with a finite number of isolated fixed points. However, the periodic oscillations in systems with a line of equilibria are not true self-sustained oscillations in a rigorous sense. Properties of the systems with a line of equilibria have not been realized in full by means of electronical analog experiment. Structural instability of attractors in such systems has been shown. Nevertheless, experimentally recorded trajectories trace an attractor of the mathematical model.
\begin{acknowledgements}
This work was supported by DFG in the framework of SFB 910 and by the Russian Ministry of Education and Science (project code 3.8616.2017/8.9). We are very grateful to A.V. Khokhlov, G.I. Strelkova and V.S. Anishchenko for helpful discussions.
\end{acknowledgements}


\end{document}